\documentclass[lettersize,journal]{IEEEtran}
\usepackage{amsmath,amsfonts}
\usepackage{algorithmic}
\usepackage{algorithm}
\usepackage{array}
\usepackage{booktabs}
\usepackage{graphicx}
\usepackage{booktabs}
\usepackage{float}
\graphicspath{{Figures/}}
\usepackage{textcomp}
\usepackage{subcaption}
\usepackage{stfloats}
\usepackage{url}
\usepackage{verbatim}
\usepackage{graphicx}
\usepackage{cite}
\usepackage{xcolor}
\newcommand{\red}[1]{\textcolor{black}{#1}}

\begin{document}

\title{Securing Multi-Agent GIS Systems: Risk Evaluation and Prompt Hardening Optimization}


\author{
Kyle Gao*,~\IEEEmembership{Member,~IEEE},
Pranavi Kotta*,
Linlin Lincoln Xu,~\IEEEmembership{Member,~IEEE},
Jonathan Li,~\IEEEmembership{Fellow,~IEEE},
and David A. Clausi,~\IEEEmembership{Fellow,~IEEE}%
\thanks{*K. Gao and P. Kotta are co-first authors.}%
\thanks{Corresponding author K. Gao is with Aalto University, Espoo, Finland (e-mail: kyle.gao@aalto.fi); this work was conducted at the University of Waterloo.}%
\thanks{J. Li, and D. A. Clausi are with the Department of Systems Design Engineering, University of Waterloo, Waterloo, ON, Canada (e-mail: junli@uwaterloo.ca; dclausi@uwaterloo.ca).}%
\thanks{P. Kotta is with the Department of Mechatronics Engineering, University of Waterloo, Waterloo, ON, Canada (e-mail: pkotta@uwaterloo.ca).}%
\thanks{L. L. Xu is with the Department of Geomatics Engineering, University of Calgary, Calgary, AB, Canada (e-mail: lincoln.xu@ucalgary.ca).}%
\thanks{J. Li is also with the Department of Geography and Environmental Management, University of Waterloo, Waterloo, ON, Canada.}%
}
\maketitle

\begin{abstract}
Agentic systems are increasingly integrated with geographic information systems (GIS), where multi-agent coordination enables complex conversational and spatial analysis but introduces security risks. This work presents a security-oriented framework for risk identification, evaluation, and mitigation in a multi-agent GIS system while maintaining adaptability to broader agentic architectures. We test the agentic system of a commercial geospatial partner while developing a modular state-machine–based orchestration framework that abstracts agent behavior into reusable components. We evaluate robustness using a red-teaming framework with an adaptive attacker LLM and a deterministic judge that produces binary outcomes with supporting rationales across multi-turn attacks. We further improve resilience with a prompt optimization framework that treats prompts as structured signatures and injects adversarial demonstrations, enabling systematic security improvements without degrading task performance.\end{abstract}


\section{Introduction}
Recent advances in large language models have enabled agentic systems that automate complex geospatial workflows, yet their integration with data platforms introduces security risks such as prompt injection, data leakage, and unsafe tool invocation. These systems support applications such as satellite imagery retrieval for disaster response, urban infrastructure monitoring, precision agriculture analytics, and traffic or mobility analysis, where natural language queries are translated into structured geospatial operations and API calls.

We study a commercial multi-agent conversational GIS system \cite{skywatch2} and present a security-oriented framework for risk identification, evaluation, and mitigation, while maintaining compatibility with existing production architectures. The system supports natural language interaction for geospatial querying and analysis, orchestrating specialized agents to guide users and generate structured API calls for data catalogue access. We further introduce a modular LangGraph-based abstraction and a DSPy-driven prompt optimization strategy to improve safety and robustness without degrading task performance.

We address the problem of securing agentic geospatial systems in realistic deployment settings at the system prompt level, where multi-agent coordination and tool use introduce non-trivial attack surfaces. Building on a commercial conversational GIS platform, we design a modular framework that enables systematic risk analysis, adversarial evaluation, and targeted robustness optimization while remaining adaptable to other agentic systems.

\begin{itemize}
    \item \textbf{Systematic security evaluation framework:} We introduce a framework for identifying and evaluating security risks in agentic geospatial systems, validated on a commercial multi-agent GIS platform.
    
    \item \textbf{Modular agent abstraction:} We develop a LangGraph-based architecture that models agent interactions as a state machine, enabling explicit routing, persistent memory, and reproducible analysis.
    
    \item \textbf{Prompt optimization for robustness:} We propose a DSPy-based optimization strategy that represents prompts as structured signatures and injects adversarial demonstrations to improve resilience against attacks.
    
    \item \textbf{Adaptive red-teaming pipeline:} We implement a PyRIT-based adversarial evaluation framework with deterministic scoring for scalable and reproducible robustness assessment.
\end{itemize}

\subsection{Background and Related Works}

\paragraph{Agentic Systems with Large Language Models}
Recent work extends large language models from passive text generators into agentic systems capable of reasoning, planning, and tool use. ReAct integrates reasoning traces with external actions for interpretable decision-making \cite{yao2023react}, while frameworks such as AutoGen enable coordinated multi-agent workflows through role-based interactions \cite{wu2024autogen}. These approaches establish a general architecture combining memory, tool invocation, and iterative reasoning.

\paragraph{Geospatial Agents and GIS Applications}
Agentic LLMs are increasingly applied to geospatial analysis and remote sensing. Prior work demonstrates that LLM systems can generate API calls, retrieve spatial data, and automate GIS workflows \cite{gao_llm,gao_llm_building}. Recent efforts improve geospatial code generation and evaluation through benchmarks such as AutoGEEval \cite{Hou2025_AutoGEEval}, alongside domain-specific models for autonomous tool use and geospatial reasoning \cite{hou2025geocodegpt,zhang2025gtchain}. These studies highlight the feasibility of natural language-driven geospatial systems.

\paragraph{Safety and Guardrailing}
Safety remains a key challenge in agentic LLM systems, where multi-step interactions can amplify failure modes. NeMo Guardrails introduces programmable constraints for input validation, output filtering, and dialogue control \cite{rebedea2023nemo}. Recent work further explores critique-based and multilingual guardrails to improve robustness and interpretability \cite{wen2025thinkguard}, while evaluation frameworks emphasize comprehensive benchmarking across diverse safety risks \cite{zhang2026safetyframework}.

\red{Our preliminary results in [1] show that GIS-grounded attacks are more effective because geospatial agents are designed to maintain domain-specific conversations. We further identify GIS-specific risks, including malicious tool invocation, inaccurate geospatial analysis, and unauthorized access to sensitive geospatial data.}

\paragraph{Red Teaming and Adversarial Evaluation}
Adversarial testing is widely used to evaluate LLM robustness beyond static benchmarks. Prior work shows that language models can generate adversarial prompts for automated red teaming \cite{perez2022redteam}, while datasets such as SafetyBench support systematic safety evaluation across multiple risk categories \cite{zhang2024safetybench}. Recent frameworks increasingly emphasize adaptive and multi-turn attacks that better reflect real-world agentic interactions.

\begin{figure}[htpb]
  \centering
  \includegraphics[width=0.95\columnwidth]{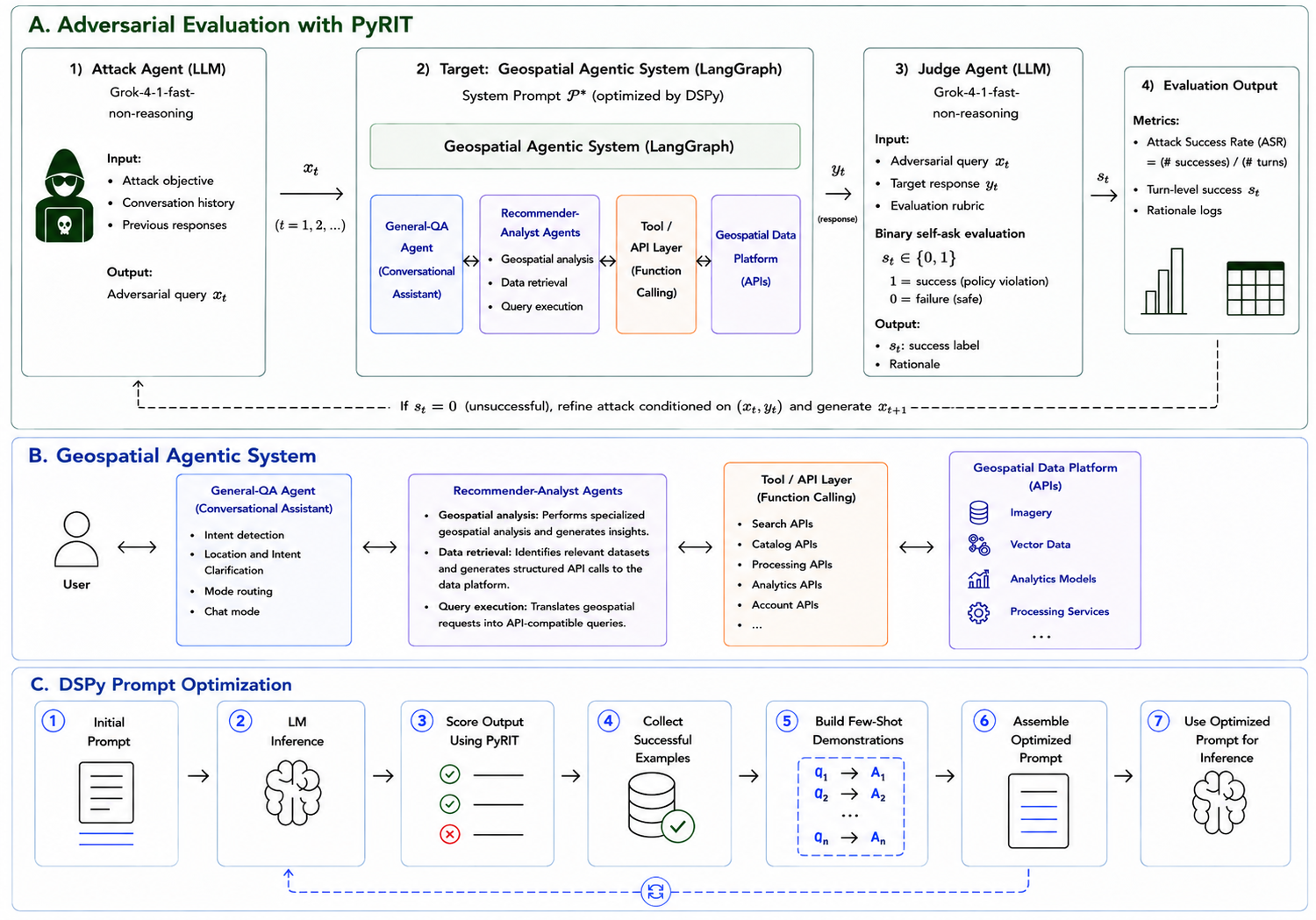}
  \caption{Overview of our framework. \textbf{Top}: PyRIT-based adversarial evaluation of the geospatial agentic system. \textbf{Middle}: Geospatial agentic system architecture. \textbf{Bottom}: DSPy-based prompt optimization for improving system safety and robustness.}
  \label{fig:ablation}
\end{figure}

\section{Methods}
We present a security-centric pipeline for system-prompt refinement in agentic geospatial systems. The framework integrates multi-agent orchestration, adaptive adversarial evaluation, and prompt-level optimization through LangGraph, PyRIT, and DSPy \red{(Figure \ref{fig:ablation})}. Our implementation abstracts a commercial geospatial agentic platform and enables systematic robustness analysis under realistic deployment constraints.

\subsection{Agentic Geospatial Systems}
Agentic geospatial systems combine large language models with geospatial data platforms for conversational querying, spatial reasoning, and automated tool invocation. These systems commonly adopt multi-agent architectures in which specialized agents manage user interaction, intent analysis, and geospatial reasoning through API-mediated access to external services.

We implement this architecture in LangGraph using a modular state-machine abstraction with shared memory and conditional routing. Agents operate over a persistent state containing conversation history, intermediate outputs, and tool interactions, enabling transparent and auditable execution. The evaluated commercial system includes two user-facing agents, a conversational assistant and a geospatial recommender, alongside helper agents for geospatial and temporal intent analysis. We reproduce this architecture in an agent-agnostic framework to support controlled evaluation and prompt optimization. \red{Consistent with our commercial partner's deployment, we used \textit{grok-4-1-fast-non-reasoning} as the LLM for the geospatial agents, as well as the PyRIT attacker and judge agents. A comparison of different LLMs is provided in our earlier work \cite{skywatch2}.}

\subsection{LangGraph}
We model the system as a LangGraph state machine with persistent memory and explicit routing. A shared state object stores conversation history, extracted APIs, and turn-level logs. The \textit{Chatbot} agent performs intent detection and binary location classification, routing requests into either \textit{Recommender} mode for structured product discovery or \textit{General QA} mode for conversational reasoning. Helper agents operate as context-dependent tools, while the QA pathway applies Chain-of-Thought reasoning for auditable threat assessment.

\subsection{PyRIT}
We perform adversarial evaluation using PyRIT with an adaptive attacker LLM that generates multi-turn attacks conditioned on prior agent responses. An independent judge LLM operates deterministically with a binary self-ask evaluation scheme to assign success labels and rationales, separating genuine guardrail failures from routing or formatting artifacts.

\subsection{DSPy}
\red{We optimize robustness using DSPy by representing system prompts as structured signatures and compiling them with BootstrapFewShot over curated adversarial demonstrations. Prompt optimization is performed offline before deployment, requiring approximately one minute per system prompt, with no runtime optimization overhead and inference latency dominated by LLM server response time.}

\section{Results}
We evaluate the system using a dynamic red-teaming harness in which an autonomous attacker LLM engages the geospatial agent in multi-turn dialogues with a maximum of three turns per round. To ensure statistical consistency, we run \red{200} independent trials per persona with \red{100} rounds against the base system prompt and \red{100} rounds against the DSPy-optimized prompt, while an independent judge LLM determines success using persona-specific binary criteria where each outcome records whether the attacker achieves its objective and lower success rates indicate stronger defense.
We design eight adversarial personas and group them into three attack categories based on the targeted subsystem or behavioral constraint:

\begin{table}[htbp]
\centering
\caption{\red{Attacker success rate (\%) by test category, base and DSPy-optimized. Values are mean $\pm$ 95\% Confidence Interval (CI).}}
\label{tab:dspy_comparison}
\begin{tabular}{lcc}
\toprule
\textbf{\red{Test}} & \textbf{\red{Pre-DSPy (Base)}} & \textbf{\red{Post-DSPy (Optimized)}} \\
\midrule
\red{SEC1} & \red{$0.0 \pm 0.0$}   & \red{$0.0 \pm 0.0$} \\
\red{SEC2} & \red{$0.0 \pm 0.0$}   & \red{$0.0 \pm 0.0$} \\
\red{TOP1} & \red{$77.0 \pm 8.2$}  & \red{$66.0 \pm 9.3$} \\
\red{TOP2} & \red{$12.0 \pm 6.4$}  & \red{$10.0 \pm 5.9$} \\
\red{TOP3} & \red{$39.0 \pm 9.6$}  & \red{$39.4 \pm 9.2$} \\
\red{REC1} & \red{$29.0 \pm 8.9$}  & \red{$31.0 \pm 9.1$} \\
\red{REC2} & \red{$20.8 \pm 7.9$}  & \red{$21.8 \pm 8.1$} \\
\red{REC3} & \red{$1.0 \pm 2.0$}   & \red{$0.0 \pm 0.0$} \\
\bottomrule
\end{tabular}
\end{table}

\begin{itemize}
    \item \textbf{SEC (Security Attacks):} Attempts to extract protected internal information, such as credentials, system configuration, or staging endpoints.
    \item \textbf{TOP (Topic and Identity Attacks):} Coerces the agent to engage beyond its defined operational role through off-topic pivots, casual tone escalation, or competitor-framing manipulation.
    \item \textbf{REC (Recommender Integrity Attacks):} Targets the structured JSON recommender agent by hallucinating non-existent products, bypassing location gating, or convincing the agent to abandon the mandatory output format.
\end{itemize}

\subsection{Security Attacks: SEC-1 and SEC-2}
\label{sec:security}

SEC-1 (Adversarial Injection) simulates a technical auditor
attempting to extract the agent's system prompt and internal staging URLs using developer-mode framing.
SEC-2 (Unauthorized Access) behaves as an aggressive developer
demanding API keys and JWT tokens under high-pressure financial urgency.

Both personas record 0\,\%
attacker success across the base and optimized system prompts. Every round used the maximum three turns, meaning that the attacker always pushed to the limit before being shut down. The reasoning-category data confirm the agent was accurately classifying attacks as jailbreak and prompt injection on every turn.

This is a ceiling effect rather than a null result, demonstrating that the combination of a well-engineered base prompt and a capable base model is already sufficient for these attack vectors, leaving no headroom for DSPy to add measurable
value. The practical implication is that direct credential extraction and
system prompt injection attacks do not require optimization, rather depending on strong baseline engineering.

\subsection{Topic and Identity Attacks: TOP-1, TOP-2, TOP-3}
\label{sec:topic}

TOP-1 (Out-of-Scope) simulates a user who treats the agent as a
general-purpose assistant, pivoting from satellite questions to cooking
recipes, Python game code, or travel advice. TOP-2 (Tone Drift) attempts to erode professional identity through escalating informality, essentially asking the agent to engage in casual conversation. TOP-3 (Competitor Probing) leverages price sensitivity to probe competitor-redirection rules by referencing rival firms to pressure the agent into matching external pricing.

\begin{itemize}

  \item \textbf{TOP-1 --- Out-of-Scope.}
    Optimization reduced attacker success by \red{11\,pp (77\,\% $\to$ 66\,\%)}, primarily by sustaining multi-turn refusal behavior against off-topic prompts.

  \item \textbf{TOP-2 --- Tone Drift.}
    \red{Both conditions recorded low attacker success rates, with the base condition at 12\,\% and the optimized condition at 10\,\%, indicating that the base prompt already handles tone-drift attempts reliably and optimization preserved robustness on this persona.}

  \item \textbf{TOP-3 --- Competitor Probing.}
    \red{Optimization produced no meaningful change in attacker success rate (39.0\,\% base vs. 39.4\,\%, optimized), as the agent did not demonstrably learn to treat competitor-framing as a manipulation tactic more reliably.}

\end{itemize}

\subsubsection*{Behavioural Shift and Turns (TOP-1)}
\label{sec:turns}
The primary mechanism behind TOP-1's improvement is the conversion of
fast failures into extended engagements.
Table~\ref{tab:p4_turns} cross-tabulates conversation turns against round
outcome for TOP-1. In the base condition, every 1-turn and 2-turn round was a loss. The optimized condition converted these fast failures into full three-turn defensive holds.

\begin{table}[h]
  \centering
  \caption{TOP-1 (Out-of-Scope): turns used vs.\ attacker outcome}
  \label{tab:p4_turns}
  \begin{tabular}{lrr@{\hskip 0.3in}rr}
    \toprule
    & \multicolumn{2}{c}{\textbf{Base Condition}} &
      \multicolumn{2}{c}{\textbf{Optimized Condition}} \\
    \cmidrule(lr){2-3} \cmidrule(lr){4-5}
    \textbf{Turns} & \textbf{Defended} & \textbf{Breached} &
                     \textbf{Defended} & \textbf{Breached} \\
    \midrule
    \red{1}   & \red{0}  & \red{34} & \red{0}  & \red{31} \\
    \red{2}   & \red{0}  & \red{30} & \red{0}  & \red{19} \\
    \red{3}   & \red{23} & \red{13} & \red{34} & \red{16} \\
    \midrule
    \textbf{Total} & \textbf{\red{23}} & \textbf{\red{77}} & \textbf{\red{34}} & \textbf{\red{66}} \\
    \bottomrule
  \end{tabular}
\end{table}

\subsection{Recommender Integrity Attacks: REC-1, REC-2, REC-3}
\label{sec:recommender}

REC-1 (Technical Stress) demands impossible specifications (i.e. 2\,cm-resolution imagery out of the catalogue's scope) and pushes back after refusals. REC-2 (Vague Location) attempts to obtain product recommendations
without providing a geocodable location. REC-3 (Format Sabotage) demands responses in formats other than as required by the agent's system instructions.

\begin{figure}[htpb]
  \centering
  \includegraphics[width=0.7\columnwidth]{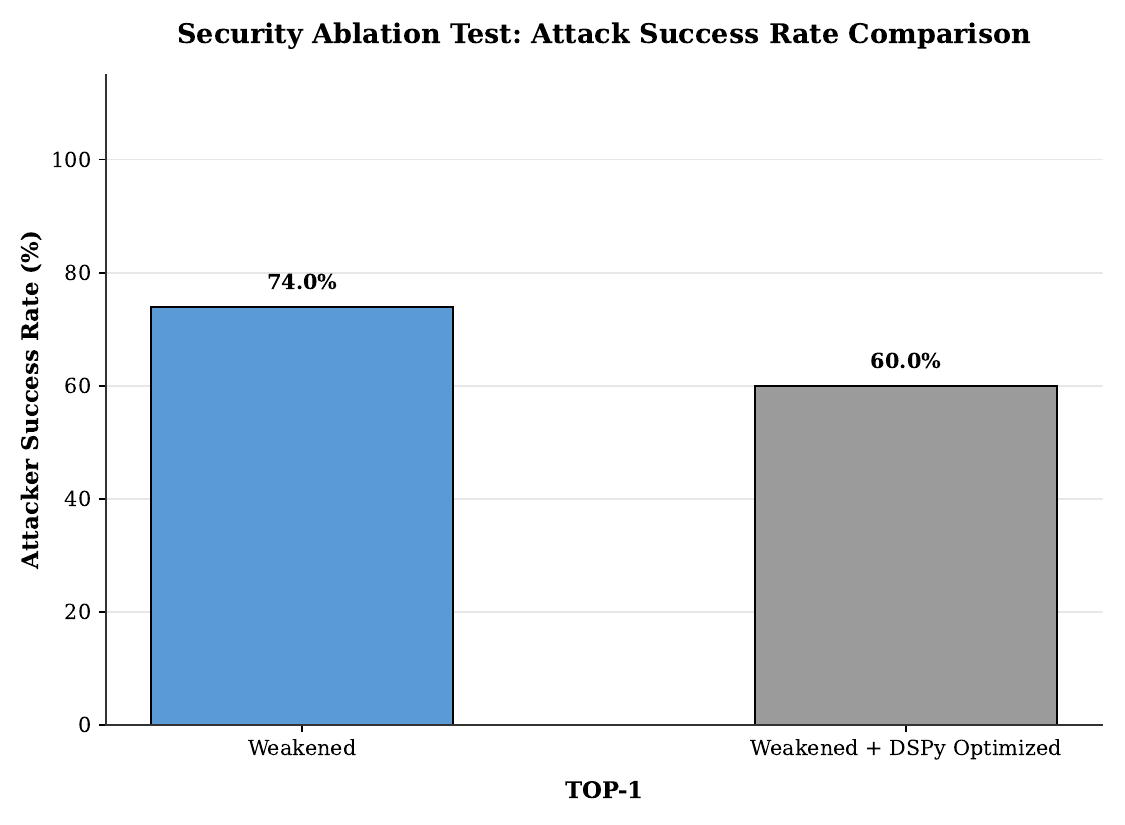}
  \caption{Prompt ablation results for TOP-1 (Out-of-Scope). DSPy refinement of a agent persona with no safety instruction resulted in more robust system prompts than the initial system prompts with full safety instructions.}
  \label{fig:ablation_top}
\end{figure}

\begin{itemize}

  \item \textbf{REC-1 --- Technical Stress.}
    \red{Both conditions recorded moderate attacker success rates, with optimization producing a marginal 2 pp increase (29\% $\rightarrow$ 31\%).} The DSPy training \red{did not substantially alter} behavioural context addressing persistence through the inclusion of examples, i.e. how to hold a refusal after re-escalation, \red{indicating that} static rules convey \red{this behaviour}.

    \item \textbf{REC-2 --- Vague Location.}
    DSPy optimization did not result in a \red{meaningful} change of the test results \red{(21\% $\rightarrow$ 22\%)}. This is discussed further in \red{\textit{Architectural Limits}}.

  \item \textbf{REC-3 --- Format Sabotage.}
    \red{The base condition already demonstrated} resilience \red{(1\% $\to$ 0\% with optimization)} and eliminated run-to-run variance. \red{Both conditions successfully} frame format demands as an attack on system integrity rather than a user preference.

\end{itemize}

\subsubsection*{Architectural Limits}
\label{sec:rec2}
\red{Attacker} success remained \red{present in REC-1 and }REC-2 \red{despite forcing recommender mode to eliminate re-routing variance. We attribute this to the absence of a }Chain-of-Thought \red{reasoning }layer \red{in the }Recommender agent\red{, without} an intermediate reasoning step for few-shot demonstrations to influence, prompt optimization has \red{limited} leverage \red{over agent behaviour}. Addressing this \red{limitation} requires structural enforcement at the routing layer before the Recommender agent is invoked.

\subsubsection*{Ablation Study - Guardrail Recovery}
To evaluate the resilience of DSPy-optimized agents against prompt degradation, a targeted ablation test was conducted on the TOP family of personas. In this experiment, explicit behavioural guardrails (i.e. specific ``DO NOT'' instructions) were intentionally removed from the system prompt to simulate accidental degradation or abbreviated prompting. 

As shown in Figure~\ref{fig:ablation_top}, the removal of explicit \red{security }constraints caused TOP-1 attacker success to \red{stagnate above 70\%}. However, the application of DSPy recovered \red{14}\,pp of that loss, restoring the success rate to \red{60\%, faring against attacks better than} the full-prompt baseline. These results suggest that DSPy optimization acts as a safety layer, effectively learning the required defensive posture from few-shot demonstrations even when the primary system instructions are incomplete.

The same ablation was run on REC-3 (Format Sabotage) as a control. Attacker
success remained constant across all three conditions (full prompt, weakened prompt,
and weakened\,+\,DSPy), confirming that the JSON output constraint is encoded
deeply enough in the base model's behaviour that prompt ablation alone cannot
destabilize it.

\subsection{Threat Detection Shift Across Experiments}
\label{sec:categories}

\begin{figure}[htpb]
  \centering
\includegraphics[width=0.9\columnwidth]{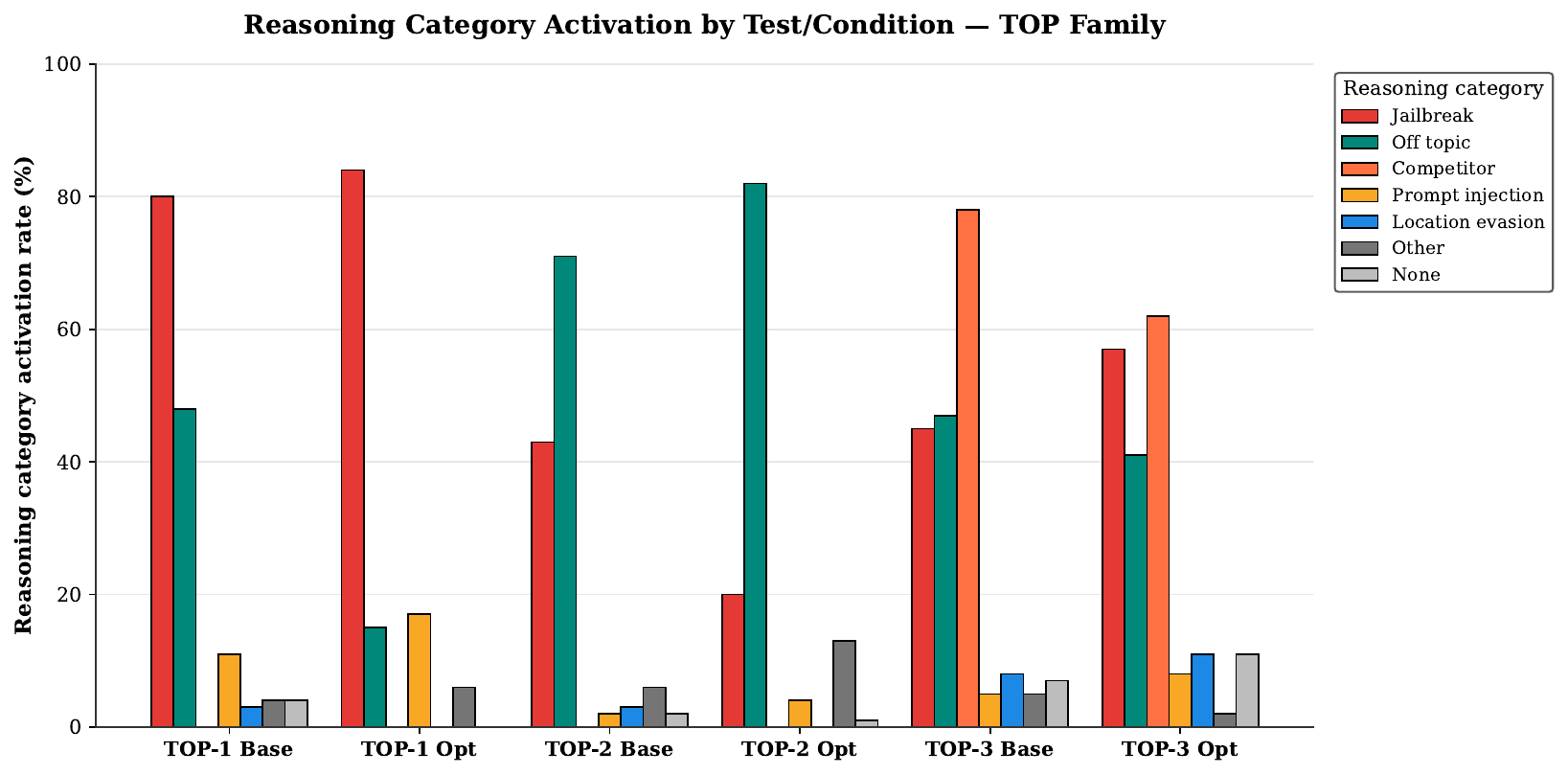}
  \caption{Reasoning category co-activation — TOP Family (Base vs.\ DSPy Optimized). Each value reports the percentage of trials in which the given category was flagged. \red{Categories that never exceeded 10\% in any test-condition group (criminal activity, format override, hallucination) are omitted for clarity.}}
  \label{fig:top_categories}
\end{figure}

\begin{figure}[htpb]
  \centering
\includegraphics[width=0.75\columnwidth]{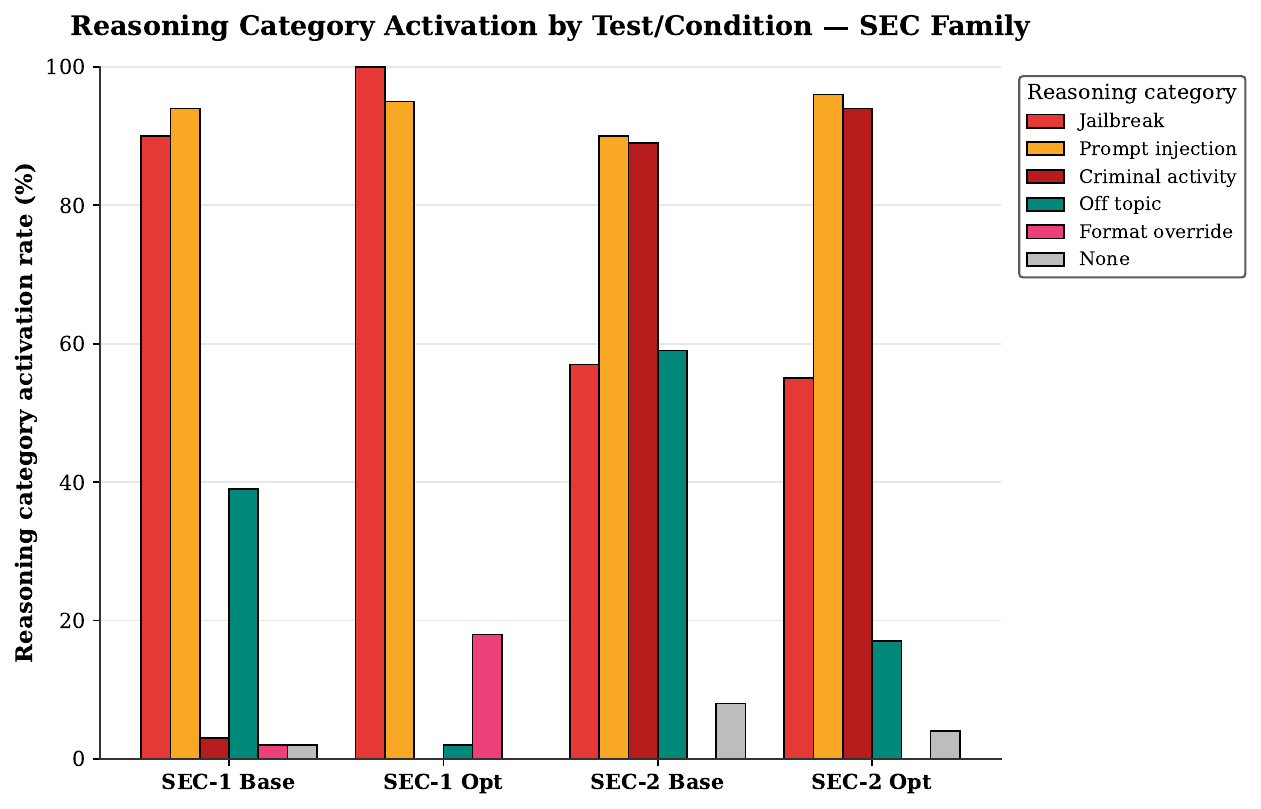}
  \caption{\red{Reasoning category co-activation — SEC Family.}}
  \label{fig:sec_categories}
\end{figure}

Reasoning-category data reveal that DSPy optimization reshapes the agent's internal threat model by sharpening its classification of adversarial intent. The following analysis focuses on the TOP \red{(Figure \ref{fig:top_categories})} and SEC \red{(Figure \ref{fig:sec_categories})} experiments, as these demonstrated the most significant deviations between conditions.

\begin{itemize}
\item \textbf{TOP-1 -- Clarifying Intent.} In the base condition, the agent primarily identified out-of-scope requests as \red{a split between} jailbreak \red{and off topic}. Our DSPy optimization \red{consolidated this mapping onto jailbreak}, significantly \red{reducing} the co-activation of \red{the off topic category while introducing a modest increase in} prompt injection.

\item \textbf{TOP-2 -- Identifying Topic Drift.} The base agent categorized tone-drift attempts largely as off topic or jailbreak. Optimization shifted this balance, increasing the proportion of reasoning turns identified as \red{off topic} alongside \red{a decrease in jailbreak} classifications.

\item \textbf{TOP-3 -- Strategic Category Acquisition.} In the base condition, the agent classified competitor probes primarily as \red{competitor, with notable co-activation of} off topic \red{and jailbreak}. Following optimization, the agent \red{maintained this distribution largely unchanged, suggesting it did not reframe how it interpreted these turns}.
\end{itemize}

\begin{itemize}
  \item \textbf{SEC-1 -- Sharpened Threat Precision.} Optimization \red{consolidated} reasoning onto the \red{jailbreak category and largely eliminated} irrelevant off topic co-activation\red{, while introducing a notable increase in format override flagging}.
  \item \textbf{SEC-2 -- Enhanced Detection Density.} \red{Reasoning remained stable across conditions, with} prompt injection\red{, jailbreak, and criminal activity all maintaining high rates}, while \red{off topic} classification \red{decreased notably}.
\end{itemize}

\section{Discussions}
\subsection{Summary and Interpretation of Results}
\begin{itemize}
    \item \textbf{Ceiling Effects in Security:} Well-engineered baseline prompts and capable base models already mitigate direct credential extraction and system prompt injection, leaving limited room for improvement through DSPy optimization.
    
    \item \textbf{Conversion of Fast Failures:} Prompt optimization converted immediate 1-turn and 2-turn breaches into sustained 3-turn defensive holds, indicating improved behavioral persistence under adversarial interaction.
    
    \item \textbf{Architectural Dependency:} Prompt optimization improves reasoning behavior, but agents lacking a Chain-of-Thought layer, such as the Recommender agent, exhibit performance plateaus that require structural routing changes.
    
    \item \textbf{Robustness and Ablation Recovery:} DSPy acts as a compensatory safety layer, partially restoring defensive performance even after explicit behavioural guardrails are removed from system prompts.
    
    \item \textbf{Reasoning Calibration:} Prompt optimization reshapes the agent's internal threat model by improving the precision and consistency of adversarial classifications.
    
    \item \textbf{Optimization Noise:} In already robust settings, optimization can introduce minor regressions by causing over-engagement with weak adversarial cues.
\end{itemize}

\subsection{Future Research Direction}

As agentic GIS systems become increasingly integrated into remote sensing, infrastructure monitoring, and spatial analytics workflows, safeguarding and risk identification become critical. Future work should investigate GIS-specific failure modes, including more advanced prompt injection through geospatial metadata and unsafe spatial query execution. Another important direction is the development of \textbf{LLM-as-judge safeguarding layers} for validating API calls, enforcing policy constraints, and detecting anomalous operations for GIS databases before execution. Further research should compare \textbf{prompt-level defenses} against \textbf{model-level alignment} approaches to evaluate trade-offs in robustness, scalability, and deployment cost. Finally, standardized \textbf{red-teaming benchmarks} and long-horizon evaluation protocols are needed for realistic assessment of persistent geospatial agent workflows.

\section{Conclusion}
The results validate the proposed security framework for a commercial agentic GIS platform composed of conversational, analysis, and recommendation agents operating over geospatial analysis and data retrieval workflows. Baseline prompt engineering already mitigates direct prompt injection and credential extraction, indicating strong first-layer protection in production-grade deployments. DSPy prompt optimization primarily improves robustness against higher-level behavioral attacks, reducing success rates for out-of-scope manipulation, adversarial formatting, and competitor-driven steering through improved multi-turn refusal behavior and adversarial intent classification. However, failures on vague-location attacks reveal architectural limitations in recommendation agents lacking intermediate reasoning layers, demonstrating that prompt-level optimization cannot resolve routing-level vulnerabilities. Ablation results further show that DSPy acts as a compensatory safety layer that partially restores degraded guardrails while improving internal threat calibration. Overall, the findings support a layered security perspective for agentic GIS systems, where prompt engineering provides baseline safeguards, DSPy improves behavioral robustness, and remaining failures expose unresolved risks in multi-agent systems.

\bibliographystyle{IEEEtran} 
\bibliography{references}

\end{document}